\documentclass[prl,superscriptaddress,showpacs,balancelastpage]{revtex4} 

\usepackage{graphicx, natbib}
\usepackage{verbatim}
\usepackage{color}
\usepackage{amsmath}
\usepackage{amsfonts}
\usepackage{amssymb}
\usepackage[latin1]{inputenc}

\begin{document}

\title{Microtraps for neutral atoms using superconducting structures in the critical state}

\author{A. Emmert}
\affiliation{Laboratoire Kastler Brossel, CNRS, ENS, UPMC-Paris 6 -- 24 rue Lhomond, 75231 Paris cedex 05, France}
\author{A. Lupa\c scu}
\affiliation{Laboratoire Kastler Brossel, CNRS, ENS, UPMC-Paris 6 -- 24 rue Lhomond, 75231 Paris cedex 05, France}
\affiliation{Institute for Quantum Computing, University of Waterloo, 200 University Av. W, Waterloo, ON, N2L 3G1, Canada}
\author{M. Brune}
\author{J.-M. Raimond}
\affiliation{Laboratoire Kastler Brossel, CNRS, ENS, UPMC-Paris 6 -- 24 rue Lhomond, 75231 Paris cedex 05, France}
\author{S. Haroche}
\affiliation{Laboratoire Kastler Brossel, CNRS, ENS, UPMC-Paris 6 -- 24 rue Lhomond, 75231 Paris cedex 05, France}
\affiliation{Coll\`{e}ge de France, 11 place Marcelin Berthelot, 75231 Paris
Cedex 05, France}
\author{G. Nogues}
\email{gilles.nogues@lkb.ens.fr}
\affiliation{Laboratoire Kastler Brossel, CNRS, ENS, UPMC-Paris 6 -- 24 rue Lhomond, 75231 Paris cedex 05, France}

\date{\today{}}

\begin{abstract}
Recently demonstrated superconducting atom-chips provide a platform for trapping atoms and coupling them to solid-state quantum systems. Controlling these devices requires a full understanding of the supercurrent distribution in the trapping structures. For type-II superconductors, this distribution is hysteretic in the critical state due to the partial penetration of the magnetic field in the thin superconducting film through pinned vortices. We report here an experimental observation of this memory effect. Our results are in good agreement with the predictions of the Bean model of the critical state without adjustable parameters. The memory effect allows to write and store permanent currents in micron-sized superconducting structures and paves the way towards new types of engineered trapping potentials.
\end{abstract}


\maketitle

Atom-chips trap ultracold neutral atoms in the magnetic field of micron-sized current-carrying wires or ferromagnetic films~\cite{fortagh_2007_RevAtomChips}. High field gradients are obtained at short distances from on-chip microfabricated structures. The resulting strong confinement considerably simplifies the production of Bose-Einstein condensates (BEC)~\cite{haensel_2001_1,ott_2001_becMicroTrap.pdf}. Moreover, complex potentials can be designed and exploited for the realization of atomic conveyor belts~\cite{hansel_2001_mcb}, beam-splitters~\cite{shin_2005_splitBEC}, and miniaturized atomic clocks~\cite{treutlein_2004_1}.

Recent experiments~\cite{nirrengarten_2006_1, mukai_2007_PersCurrChip, cano_2008_MeissnerEffectMicrotraps} have demonstrated that the normal metal wires can be replaced by superconducting materials. The trapping time is expected to be extremely long  close to superconductors~\cite{SCHEEL_2005, skagerstam_2006_SpinDecohSuperchips, hohenester_2007_sfl, nogues_2009p_VorticesLifetime}, due to a considerable reduction of the Johnson-Nyquist noise, which is the dominant source of losses for normal metal atom-chips. Moreover, the cryogenic environment of superconducting chips makes it possible to couple cold ground state or Rydberg atoms with solid-state quantum devices, such as linear coplanar microwave cavities~\cite{petrosyan_2009_AtomicEnsemblesAndQubits} or Josephson devices~\cite{singh_2009_SquidIntBec}. These hybrid systems will open new avenues for fundamental research and have potential applications in quantum information processing~\cite{hyafil_2004_2}.

Cold atoms on chips are also sensitive probes of the magnetic trapping potential and, hence, of the current distribution in the wires. Tiny current fluctuations caused by edge imperfections can explain atomic cloud fragmentation close to the chip \cite{Esteve_2004}. Sensitive imaging of the magnetic field by a BEC  has been used for probing current distributions in metallic films \cite{WILDERMUTH_2005}. Current distributions in superconducting films are expected to be dramatically different from those in normal metals. Cano \textit{et al.} have recently shown that trapping frequencies decrease significantly  close to a cylindrical 
niobium wire due to the screening currents expelling the magnetic field from the superconductor (Meissner effect)~\cite{cano_2008_MeissnerEffectMicrotraps}. However, this Meissner regime does not apply to superconducting thin films, where large enough magnetic fields partially penetrate the film through vortices. The Bean model of the critical state~\cite{bean_1962_MagHardSuperc, bean_1964_BeanModel, Henkel_2009} provides a theoretical description of the current distribution in this mixed phase. It predicts hysteresis for the film magnetization, which cannot be described by the Meissner thermodynamic phase.

We present here an experimental evidence of this hysteretic behavior. We bring ultracold atoms close to a superconducting thin strip. The atomic cloud follows trajectories which reveal different magnetic field maps depending on the magnetization history of the superconductor. Our observations are in good agreement with numerical calculations based on the Bean model~\cite{prigozhin_1996_BeanVar}, with no adjustable parameter. These results show that permanent currents present in superconducting films can be understood quantitatively and controlled at the micrometer scale, opening the way to new trap geometries.

\begin{figure}[t]
\begin{center}
\includegraphics[width=8.5cm]{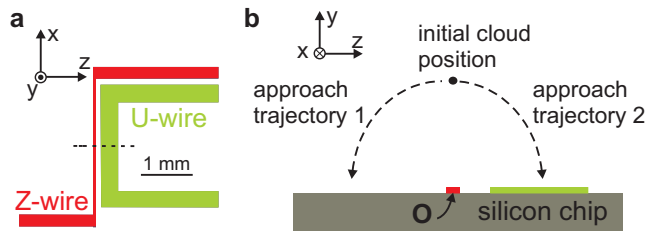}
\end{center}
\caption{\label{fig1}(a) Central part of the chip, showing the U- and Z-wires. (b) Cross-section of the chip along the dashed line in (a). The initial trap position and the two approach trajectories are also represented. The position of the cloud is measured with respect to the origin $O$ at the center of the Z-wire. }
\end{figure}

The layout of our atom chip is shown in Fig.~\ref{fig1}a and is described in detail in Ref.~\cite{nirrengarten_2006_1}. The chip is based on a 360~$\mu$m thick silicon substrate covered by a 500~nm thick layer of silicon oxide. The structures used for trapping are a U-shaped wire and a Z-shaped wire. These wires are obtained by sputtering a 1~$\mu$m niobium layer (transition temperature $T_c=9.2$~K), which is then patterned using standard optical lithography. The central regions of the U- and Z-wires have widths of 300~$\mu$m and 40~$\mu$m respectively. The measured critical current of the Z-wire is 1.8~A; that of the U-wire is larger than 5~A. Following wire patterning, a 1.5~$\mu$m planarization layer is added as a base for a 200~nm thick gold layer. It acts as a mirror used for the mirror magneto-optical trap (MOT), which traps and cools the $^{87}$Rb atoms before loading in the chip magnetic trap. The chip is vertically mounted on the cold finger of a custom-made cryostat, operated at a temperature of 4.2~K.

The sequence for atom cooling and trapping is detailed in~\cite{nirrengarten_2006_1,roux_2008_1}. The U-wire is used in combination with an uniform bias field to create the quadrupolar magnetic field required for the operation of the mirror-MOT. After trapping and cooling in the mirror-MOT, the atoms are further cooled using optical molasses and then optically pumped to the $|5S_{1/2},F=2, m_{F}=2\rangle$ hyperfine sublevel of the ground state of $^{87}$Rb. We then abruptly switch on the trapping potential generated by a current $I_{Z}$ in the Z-wire and a bias magnetic field $(B_x,B_y,B_z)$. The cloud is then compressed and evaporatively cooled. We finally adiabatically set the Z-wire current to $I_Z=1.34$~A and the bias magnetic field to $(-3.0,0,9.4)$~G. The $\approx 1.4\cdot10^5$ atoms are then at a temperature of $30(5)\ \mu$K and at a distance of about 250~$\mu$m above the Z-wire. 

\begin{figure}[t]
\begin{center}
\includegraphics[width=8.5cm]{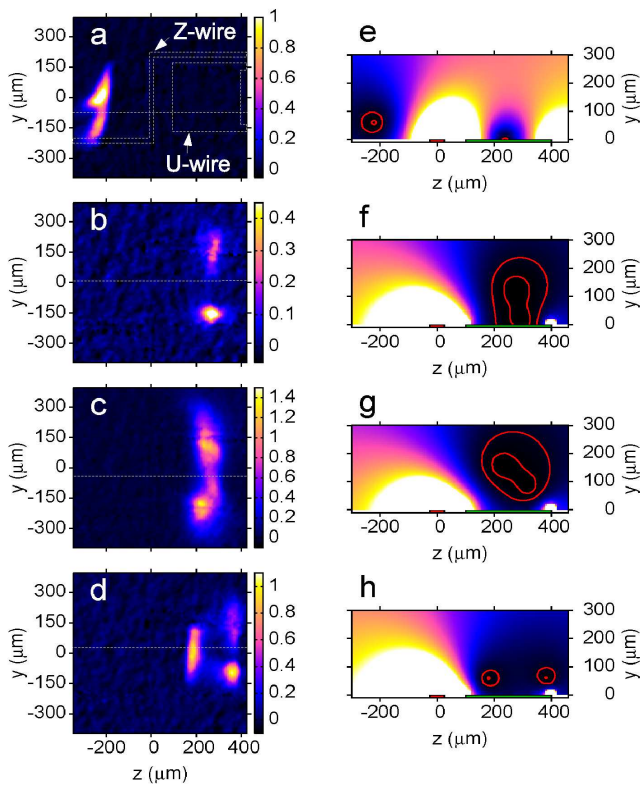}
\end{center}
\caption{\label{fig2} Absorption images of the atomic cloud for different approaches to the chip surface. The final values of the external bias field are given in the text. The color scales indicate the optical density. The dotted horizontal lines separate the direct (top) and reflected (bottom) images of the atomic cloud. (a) Approach along trajectory 1 in Fig.~\ref{fig1}b. (b--d) Approach along trajectory 2 for $B_{y,t}=$0~G (b) and $B_{y,t}=$-3~G (c--d). (e--h) 2D-plot of the calculated trapping potential for the same conditions as in Figs.~(a--d). 
The two solid lines correspond to equipotential lines at 5 and 30~$\mu$K respectively. The red (left) and green (right) rectangles above the abscissa represent the extension of the Z-wire and the U-wire respectively along the $z$-axis.}
\end{figure}

The final approach to the chip is controlled by linearly sweeping, during 400~ms, $B_y$ and $B_z$ towards the final values $(B_{y,f},B_{z,f})$ while keeping $B_x=-3~$G constant. After each sequence, an absorption image of the atomic cloud is taken by using a probe beam sent along an axis in the $xy$-plane, at an angle of 11$^{\circ}$ with respect to the $x$-axis. The absorption image shows the cloud and its reflection in the on-chip gold mirror (see Fig.~\ref{fig2}). A gaussian fit of the direct and reflected images provides the vertical position $z_{c}$ and the distance $y_{c}$ to the chip~\cite{roux_2008_1}.

We first obtain a reference trajectory by approaching the atoms to the surface while keeping a large distance from superconducting wires~\cite{emmert_2009_lifeatomscoldchip}. In order to image the cloud trajectory, we repeat 61 times the full sequence varying linearly $(B_{y,f},B_{z,f})$ from (0,9.4)~G to (14.1,0)~G. The cloud follows the nearly circular trajectory labeled~1 in Fig.~\ref{fig1}b. The atom--Z-wire distance remains always larger than 240~$\mu$m, while the distance $y_{c}$ to the chip surface decreases.  Fig.~\ref{fig2}a presents an absorption image for $(B_{y,f},B_{z,f})= (10.1,\ 2.7)$~G. 
We observe the cloud (on top) and its reflection. The cloud has a diameter of 26(3)~$\mu$m and a length along the $x$-axis of 280(3)~$\mu$m. This finite length accounts for the partial overlap of the direct and reflected images. Moreover, due to the asymmetry of the Z-wire with respect to the $yz$-plane, the cloud is slightly distorted and displaced by -380(5)~$\mu$m in the $x$-direction. 
The distance of the atoms to the surface is measured to be $y_c=81.7(6)~\mu$m. For $y_c\lesssim~50\ \mu$m the atoms are rapidly lost by adsorption on the surface. 

We then study the influence of the U-wire by approaching the atoms along the trajectory labeled 2 in Fig.~1b ($(B_{y,f},B_{z,f})$ vary from (0,9.4)~G to (-14.1,0)~G). We compare different magnetization histories for the atom-chip. They correspond to cooling down the superconducting wires through the transition temperature $T_c$ with different applied perpendicular fields $B_{y,t}$. Fig.~\ref{fig2}b corresponds to $B_{y,t}$=0~G and $(B_{y,f},B_{z,f})=$(-6.6,5.0)~G. The direct image of the cloud (on top) is less visible than the reflected one because of diffusion of the probe beam on the chip wires. For this magnetization history the atoms are lost when the approach-distance $y_c$ is lower than $\approx 140~\mu$m. This minimum distance is significantly larger than that of trajectory~1.  

For Figs.~\ref{fig2}c--d, $B_{y,t}$ is equal to -3.0~G and $(B_{y,f},B_{z,f})=$(-5.4,5.8)~G and (-8.7,3.6)~G respectively. We observe that the cloud splits in two parts in between Figs.~2c and~2d (note that in Fig.~2d the direct and reflected images of the left cloud overlap). This splitting is radically different from the early cloud disappearance observed with the other magnetization history (Fig.~\ref{fig2}b). Movies of the complete cloud trajectories are available in \cite{EPAPS}. These approaches to the chip surface clearly exhibit the influence of permanent hysteretic currents in the U-wire on the trapping potential. Note that the cloud configuration is found to be quite stable, leading to reproducible images over a few days, in spite of repeated changes of applied currents and fields. The permanent current distribution is only reset by a transition to the normal state.

The current distribution in a type-II superconductor is known to be well described by the Bean critical-state model~\cite{bean_1962_MagHardSuperc,bean_1964_BeanModel,fitz_1969_beanexp}. It considers  quantities averaged over lateral dimensions larger than the intervortex distances and the thickness of the sample. It thus describes correctly a situation in which the distance between the atomic cloud and the U-wire is much larger than these length scales. The main assumption of the model for a thin film is that the surface current density $\mathbf{K}$ has a modulus always smaller than a maximum value $K_C$. When the current tends to exceed locally $K_C$, vortices rearrange, leading to a reconfiguration so that $|\mathbf{K}| \le K_C$. The Bean model relies thus on a single parameter, $K_C$, which in our case can be directly inferred from the measured Z-wire critical current: $K_C=45~$mA/$\mu$m.

There are analytical solutions of the model in simple geometries~\cite{brandt_1998_diskscylaxial, brandt_1993_typeII}, which do not apply to our situation. In order to calculate the current distribution, we use thus a numerical procedure based on a variational formulation of the model by Prigozhin~\cite{prigozhin_1996_BeanVar}. It applies to two-dimensional geometries. We assume that the central parts of the U- and Z-wires extend to infinity along the $x$-direction. This assumption is reasonable, since the length of the central parts of the wires (2~mm) is much larger the atom-chip distance. Moreover, the parts of the wires parallel to the $z$-axis (see Fig.~\ref{fig1}a) generate a field oriented mainly in the $x$-direction, which does not affect to first order the trap position in the $yz$-plane.

With these assumptions, the surface current density $\mathbf{K}$ is oriented along the $x$-axis. The U- and Z-wires are discretized in elements with a 3~$\mu$m width. Applying this discretization to a single wire, we recover accurately the available analytical result \cite{brandt_1993_typeII}. The complete numerical simulation takes into account the variations of the applied currents in the Z- and U-wires as well as those of the bias field during the whole experimental sequence. From the current distributions, we compute the magnetic field in the $yz$-plane. Figs.~2e-h present the map of the trap potential, proportional to the magnetic field amplitude, for the configuration of Figs.~2a-d respectively. The thick solid lines represent the 5 and 30~$\mu$K energy equipotentials. The predicted centers and extensions of the clouds are in good agreement with the observations. In the case of Fig.~2f, a shallow minimum is located on the surface. When the atoms are brought slightly closer to the chip, the potential barrier between the two wells lowers and the atoms rapidly escape towards the surface. In the case of Fig.~2h, the second minimum is always above the chip surface. As the bias field is varied, the two potential wells get populated.

\begin{figure}
\begin{center}
\includegraphics[width=6.5cm]{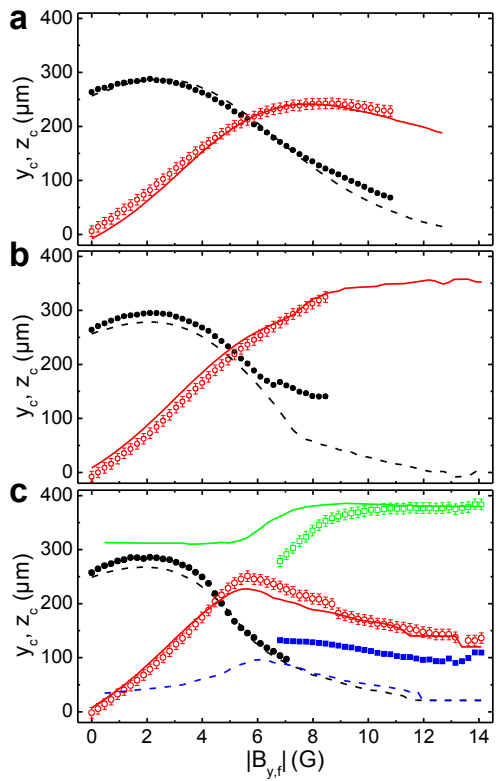}
\end{center}
\caption{\label{fig3}  Cloud center coordinates $(y_c,z_c)$ as a function of the amplitude of the external bias field $|B_{y,f}|$ for (a) approach trajectory 1, and (b-c) approach trajectory 2 for $B_{y,t}=$0 and $-3$~G respectively. Full (open) symbols and dashed (solid) lines correspond to the coordinate $y_c$ ($z_c$) inferred from the experimental data and from the theoretical Bean model respectively (the noise in the theoretical data is of numerical origin). For experimental data, it is sometimes impossible to extract one of the two or both coordinates from the image. This is the case when the cloud and its reflection on the chip surface are indistinguishable or when the signal to noise ratio becomes too small due to low atom number.}
\end{figure}

For a more quantitative analysis, Fig.~\ref{fig3} presents the comparison of the coordinates of the cloud centers $(y_c,z_c)$ as a function of the final bias field absolute value $|B_{y,f}|$ together with the numerical prediction. The approach trajectory starts at $|B_{y,f}|=$0~G. Fig.~\ref{fig3}a corresponds to the reference trajectory 1. Fig.~\ref{fig3}b-c correspond to trajectory 2 after a transition of the superconducting layer in a bias field $B_{y,t}=$0 and -3~G respectively.  The experimental data (points) are in good agreement with the predictions (lines), with no adjustable parameter. The comparison of Figs.~3a and~3b shows that the presence of the superconducting layer prevents the atoms from getting close to the surface. In Fig.~3b, above $|B_{y,f}|$=6~G, we observe that the experimental distance to the surface $y_c$ is slightly larger than the theoretical prediction. 

In the case of Fig. 3c, theory predicts the existence of two separate potential wells above the chip surface for all values of $B_{y,f}$. For $|B_{y,f}| < 6$~G, atoms are confined to the potential minimum which is initially above the Z-wire (open and full circles). The two wells nearly merge around $|B_{y,f}|$=6~G. After merging, both wells are populated as observed in Fig.~2d. The coordinates of the second well are represented by open and full squares. Note that, for $|B_{y,f}|>7$~G, we cannot measure $y_c$ for the left cloud as direct and reflected images then merge (see Fig.~2d). As in Fig.~\ref{fig3}b, the distance to the surface $y_c$ of the second well is slightly underestimated by theory. We attribute these discrepancies to the assumption of infinitely long U- and Z- wires along the $x$-direction. For large values of $B_{y,f}$, we measure that the cloud moves along $x$ by up to 0.5~mm and gets closer to the bends of these wires. A three-dimensional calculation of the current distribution and trapping potential would be better suited to this situation.

These results show that hysteretic permanent currents are an essential feature of type-II superconducting atom-chips. The trapping potential is quite different from that of normal structures. The experiments are in good agreement with numerical calculations based on the Bean model without adjustable parameter. These phenomena can be used for new programmable trapping geometries. A superconducting current-carrying wire guiding atoms in a one-dimensional potential well can be placed between two rows of square niobium patches, whose size and spacing are in the $\mu$m range. If these patches become superconducting in a bias field $B_{y,t}$, permanent currents create a modulation of the trapping potential. These currents can be cancelled by a selective heating of the patches. Switching is fast and requires low energy when the current and temperature are close to the critical values \cite{photondetect}. It can be triggered by a weak laser pulse or by a resistive element integrated in the chip under the patch. Erasing the magnetization of a selected set of patches realizes programmable superlattices or disordered potentials on a time scale much faster than the typical atomic oscillation period ($\sim 100\ \mu$s). This device opens interesting avenues for the study of atomic transport and localization. In a more complex setting, atoms trapped in a plane parallel and close to the chip by a laser standing wave can be laterally confined by the potential created by an array of magnetized superconducting patches. This programmable confinement opens the way to experiments on two-dimensional transport in complex geometries. These perspectives are also relevant for hybrid quantum systems~\cite{petrosyan_2009_AtomicEnsemblesAndQubits, singh_2009_SquidIntBec}, with atomic systems interacting with solid-state quantum structures.

\begin{acknowledgments}
We acknowledge useful discussions with L. Prigozhin, B. Pla\c{c}ais and P. Mathieu. We acknowledge support of the EU (CONQUEST and SCALA projects), of the Japan Science and Technology corporation (International Cooperative Research Project: ``Quantum Entanglement'') and of the R\'{e}gion Ile de France (IFRAF and Cnano consortiums). A.L. acknowledges support from the EU (Marie Curie fellowship) and NSERC.
\end{acknowledgments}



\end{document}